\documentclass[aps,prd,showpacs,amssymb,twocolumn,nofootinbib]
{revtex4}

\usepackage{amssymb,amsmath}
\usepackage{amsfonts}
\usepackage{mathrsfs}
\usepackage{txfonts}
\usepackage{marvosym}

\def\be{\begin{equation}}
        \def\ee{\end{equation}}
        \def\ba{\begin{eqnarray}}
        \def\ea{\end{eqnarray}}

\def\G{\Gamma}        
\def\H{{\cal H}}      
\def\fock{{\mathscr{F}}} 
\def\R{\mathbb{R}}     
\def\N{\mathbb{N}}     
\def\Z{\mathbb{Z}}     
\def\hd{\mathscr{H}}   
\def\dlapse{N_{_{_{\!\!\!\!\!\!\sim}}\;}} 
\def\dfun{F_{_{_{\!\!\!\!\!\!\sim}}\;}} 

\begin{document}

\title{Feasibility of a Unitary Quantum Dynamics
in the Gowdy $T^{3}$ Cosmological Model}
\author{Jer\'onimo Cortez}
\thanks{jacq@iem.cfmac.csic.es}
\author{Guillermo A. Mena Marug\'an}
\thanks{mena@iem.cfmac.csic.es}
\affiliation{Instituto de Estructura de la Materia. Centro de
F\'\i sica Miguel A. Catal\'an, CSIC.\\ Serrano 121, 28006 Madrid,
Spain.\\ }

\begin{abstract}
It has been pointed out that it is impossible to obtain a unitary
implementation of the dynamics for the polarized Gowdy $T^{3}$
cosmologies in an otherwise satisfactory, nonperturbative canonical
quantization proposed for these spacetimes. By introducing suitable
techniques to deal with deparametrized models in cosmology that
possess an explicit time dependence (as it is the case for the
toroidal Gowdy model), we present in this paper a detailed analysis
about the roots of this failure of unitarity. We investigate the
impediments to a unitary implementation of the evolution by
considering modifications to the dynamics. These modifications may
be regarded as perturbations. We show in a precise manner why and
where unitary implementability fails in our system, and prove that
the obstructions are extremely sensitive to modifications in the
Hamiltonian that dictates the time evolution of the symmetry-reduced
model. We are able to characterize to a certain extent how far the
model is from unitarity. Moreover, we demonstrate that the dynamics
can actually be approximated as much as one wants by means of
unitary transformations.
\end{abstract}


\maketitle


\section{Introduction}

Symmetry-reduced models have been used over the past 30 years as an
appropriate arena to test strategies aimed for the quantization of
the full theory of gravity within the canonical approach, as well as
toy models which can provide us with insights about the kind of
phenomena that should be expected in quantum general relativity.
Most of the examples of symmetry-reduced models studied so far are
minisuperspace models \cite{misner}, namely, simple systems where
the reduction leaves only a finite number of physical degrees of
freedom. There is another class of models which has more interest
inasmuch as they retain the field complexity of general relativity.
These are the so-called midisuperspace models (for a recent review
see Ref. \cite{torre-ijtp}), which after reduction possess an
infinite number of degrees of freedom. Thus, their quantization
would lead to a true quantum field theory.

Within this class, and together with the Einstein-Rosen waves
\cite{many}, the model that has deserved more attention lately is
the Gowdy $T^3$ cosmological model
\cite{guillermo-gowdy,pierri,ccq-t3,torre-prd,pierri2,toadd}. This
model was introduced by Gowdy during the seventies in a systematic
search for all spacetimes with two commuting spacelike Killing
vector fields and compact spatial hypersurfaces \cite{gowdy}. Apart
from the $T^3$ topology of a three-torus, the other two possible
spatial topologies for the Gowdy spacetimes are the three-handle
$S^1 \times S^2$ and the three-sphere $S^3$ (or the topology of a
manifold covered by one of the above). The interest in the Gowdy
$T^3$ model can thus be easily understood, since it provides the
simplest of all the inhomogeneous, empty, spatially closed
cosmological systems. The genuine field-theory character of this
model and its possible applications in cosmology make it a natural
candidate to study fundamental questions about canonical quantum
gravity and quantum field theory in curved spacetimes.

Its quantization was already considered in the seventies
\cite{varios}, and revisited both in the eighties
\cite{berger,husain,husain-smolin} and more recently
\cite{guillermo-gowdy,pierri,ccq-t3,torre-prd,pierri2}. The first
preliminary attempts to define a quantum theory and extract physics
from the Gowdy $T^3$ model \cite{varios,berger} were followed by
more detailed analysis \cite{guillermo-gowdy,husain-smolin} that
discussed the nonperturbative quantization of the system employing
the Ashtekar formulation of Lorentzian general relativity
\cite{ash-new-var}. Considerable progress has been achieved lately
in defining a complete quantization of the (sub-)model with linear
polarization, in which both Killing vectors are hypersurface
orthogonal \cite{pierri}. The proposed quantization is based on the
fact that the polarized model can be equivalently treated as $2+1$
gravity coupled to a massless scalar field, defined on a manifold
whose topology is $T^2 \times \R$.

One important aspect in the study of quantum cosmological models is
their dynamical evolution. For the polarized Gowdy $T^3$ model with
the particular quantization performed in Ref. \cite{pierri}, it has
been recently shown that the dynamics is not implementable at the
quantum level as a unitary transformation \cite{ccq-t3,torre-prd}.
From the point of view of canonical quantum gravity, this result
does not represent a serious drawback for the simple reason that
(owing to the compact nature of the spatial slices) time evolution
is pure gauge in the Hamiltonian description. Hence, there is no
time evolution and no dynamics. The system is endowed with a
fictitious dynamics via a ``deparametrization'' procedure, and there
is no apparent reason to select a preferred deparametrization.

Nevertheless, if one accepts that unitary evolution is a key
ingredient in conventional (field) quantum theory, necessary in
order to pose physically meaningful questions for issues like those
concerning the initial singularity in cosmology, the lack of a
unitary time evolution is a drawback for the kind of quantization
put forward in Ref. \cite{pierri}. From that quantum theory, one
would not be able to extract predictions for different instants of
time, because probability is not conserved. In this sense, the
quantization is not fully consistent \cite{ccq-t3}. Thus,
restoration of unitarity in the evolution seems a fundamental issue
in order to achieve a satisfactory quantization of the polarized
Gowdy $T^3$ model. A rigorous quantization along these lines will
provide us with a specific example of midisuperspace that can be
very helpful, as a point of reference for comparisons, for a future
quantization starting from loop quantum gravity (where impressive
progress has been made, but exclusively for minisuperspaces
\cite{bojowald-ash-lew}) or for implementations of the ``consistent
discretization'' approach \cite{gam-pul-1} (which has recently been
applied to the Gowdy model \cite{gam-pul-2}).

In this work, we will explore the reasons behind the failure in the
unitary implementability of the dynamics as a first step in
proposing solutions to it or introducing alternative descriptions
for the quantum evolution. Although the lack of unitarity certainly
follows from the absence of square summability for the antilinear
part of the Bogoliubov transformation which relates the annihilation
(creation) operator at, say, the ``final'' time $t_{f}$ with the
annihilation and creation operators at an ``initial'' time $t_{i}$
\cite{ccq-t3,torre-prd}, we want to analyze in detail the roots of
this failure. We will show in a precise manner why unitary
implementability fails in our system and, in a certain sense, we
will be able to characterize how far the model is from unitarity. In
doing so, we will introduce suitable techniques to deal with
deparametrized models in cosmology that possess an explicit time
dependence, like the Gowdy one.

The structure of the paper is as follows. In Sec.
\ref{sec-calssical-theory} we obtain the reduced phase space of the
system by performing the symmetry reduction and introducing a
deparametrization procedure in order to (partially) fix the gauge
freedom\footnote{The ``q-number'' nature of the time variable which
occurs in Refs. \cite{pierri,ccq-t3} will be avoided here by
introducing a gauge condition that is slightly different to the one
imposed in those references; in this way the physical degrees of
freedom will be neatly disentangled from the time variable. This
type of mixing is also absent in Ref. \cite{torre-prd}; however, the
disentanglement is attained there thanks to the introduction of an
appropriate (partially) reduced line element since the very
beginning, rather than to its construction by gauge fixing.} and
select a Hamiltonian vector field to represent the dynamics. This
reduction and gauge-fixing process had not been presented before
starting with the spacetime metric of the model in general
relativity, although the resulting description of the vacuum Gowdy
spacetimes is essentially the same that was discussed in Refs.
\cite{pierri,ccq-t3} (except for the remark on footnote$^{1}$) and
Ref. \cite{torre-prd}. Therefore, the quantum theory on which we
will base our analysis is that constructed by Pierri \cite{pierri}
(or, equivalently, that analyzed in Ref. \cite{torre-prd}).

Sec. \ref{sec-dynamics} is divided in three parts. In the first one,
we review the dynamics of the scalar field that represents (most of)
the true degrees of freedom of the theory. In this part, a crucial
remark is that the coordinates of the covariant phase space (namely
the coefficients that determine the field in terms of an orthonormal
basis of solutions -defined essentially by the negative of the
complex structure given in Ref. \cite{pierri}-) do not evolve in
time. In these coordinates, the generator of the evolution is
obviously the zero Hamiltonian. The dynamics will be introduced
through a time-dependent map from the covariant to the canonical
phase space. In the remaining parts of the section, we will
investigate the impediments for a unitary implementation of the
evolution by considering modifications to the dynamics that may be
regarded as {\em small} perturbations. We will be able to identify
where the failure of unitarity comes from and, in the last
subsection, prove that the severity of the problem is greatly
ameliorated by the fact that small corrections to the dynamics can
be implemented in a unitary way. In addition, our analysis makes
clear that the diagonalization of the Hamiltonian performed in Ref.
\cite{ccq-t3,pierri2} is just an instantaneous diagonalization which
ignores the change in time of the Bogoliubov coefficients. Here,
these time variations are explicit and rigorously taken into
account. Finally, the conclusions and some further comments are
presented in Sec. \ref{sec-conlusions}. One appendix is added which
contains a proof about the behavior of the coefficients employed in
the main discussion. In the following, lower case Latin indices on a
tensor will denote its purely spatial components, whereas capital
case Latin indices will be used to denote the tensor itself
(abstract index notation).


\section{The polarized Gowdy model}
\label{sec-calssical-theory}

The polarized Gowdy $T^3$ model describes globally hyperbolic
four-dimensional vacuum spacetimes, $({\cal M},g_{AB})$, with two
commuting hypersurface orthogonal spacelike Killing fields and
compact spacelike hypersurfaces homeomorphic to a three-torus. Since
global hyperbolicity implies that we can foliate $({\cal M},g_{AB})$
by Cauchy surfaces, $\Sigma_{t}$, parametrized by a global time
function $t$, then a $3+1$ decomposition is available and the line
element can be written \be \label{metric-general} ds^2 = -N^2 dt^2 +
h_{ij}\big(dx^i + N^i dt\big)\big(dx^j + N^j dt\big) \, , \ee where
we choose $t\in \R^+$, the coordinates in the sections of constant
time are $\{x^i\}:=\{x^1 = \theta , x^2 = \sigma , x^3 = \delta\}$
with $x^i\in S^1$, $N$ and $\{N^i\}$ are, respectively, the lapse
function and the components of the shift vector $N^A$, and
$\{h_{ij}\}$ are the components of the induced spatial metric
$h_{AB}$.

In addition, we will impose that $(\partial / \partial x^a)^A$
($a=2,3$) are the two spacelike Killing vector fields. Thus, the
metric must be independent of the coordinates $x^a$. Moreover,
performing a (partial) gauge fixing along the lines explained in
Ref. \cite{guillermo-montejo} (for pure gravitational plane waves)
and Ref. \cite{guillermo-gf} (for cylindrical spacetimes),
remembering that the metric functions must be periodic in $\theta$,
and using that $(\partial/\partial x^a)^A$ are commuting
hypersurface orthogonal vector fields, one gets a line element for
the reduced model of the form \be \label{metric-1-red} ds^2 = -N^2
dt^2 + h_{\theta \theta}\big[d\theta + N^\theta dt\big]^2
+\sum_{a=2}^3h_{aa}\left(dx^a\right)^2 \,. \ee

Let us consider now the following change of metric variables
$\{h_{ij}\} \mapsto \{Q^{\alpha}\}:=\{\psi , \gamma , \tau\}$,
defined by \be \label{new-var} h_{\theta \theta}=e^{\gamma - \psi}
\,, \quad h_{\sigma \sigma}=e^{- \psi}\tau^2 \, , \quad h_{\delta
\delta}=e^{\psi}\, . \ee Since this change is just a point
transformation, the momenta $P_{\alpha}$ canonically conjugate to
$Q^{\alpha}$ are \be P_{\alpha}=p^{ij}\,\frac{\partial
h_{ij}}{\partial Q^{\alpha}}\,,\ee where
$p^{ij}=\sqrt{h}(K^{ij}-Kh^{ij})$ are the momenta canonically
conjugate to $h_{ij}$, we have set the Newton constant $G$ equal to
$\pi/4$, $h$ is the determinant of the induced metric, and
$\{K^{ij}\}$ are the components of its extrinsic curvature, with
trace equal to $K$. Substituting in Eq. (\ref{metric-1-red}) our new
set of variables and introducing the densitized lapse factor
${\dlapse}:=N/\sqrt{h}$, we arrive at the line element \ba
\label{metric-new-var-1} ds^2&=&e^{\gamma - \psi}\biggl(-\tau^2
{\dlapse}^{2} dt^2+[d\theta + N^{\theta}dt]^2\biggr) \nonumber\\
&+&e^{-\psi}\biggl(\tau^2 d\sigma^2 + e^{2\psi}d\delta^2\biggr)\, ,
\ea whose Einstein-Hilbert action is then given by \ba
\label{action-v1} S&=&\int_{t_{i}}^{t_{f}}dt\oint d \theta\, \biggl[
P_{\alpha}\dot{Q}^{\alpha}- {\mathscr{H}}(Q^{\alpha},P_{\alpha})
\biggr]\nonumber\\ &=& \int_{t_{i}}^{t_{f}}dt\oint d\theta \,
\biggl[
P_{\tau}\dot{\tau}+P_{\gamma}\dot{\gamma}+P_{\psi}\dot{\psi}-
({\dlapse} \tilde{{\cal{C}}}+ N^{\theta}{\cal{C}}_{\theta})\biggr].
\ea The presence of the remaining first class constraints
$\tilde{\cal{C}}$ and ${\cal{C}}_{\theta}$ reflects the fact that
the gauge has been only partially fixed. These (densitized)
Hamiltonian constraint and momentum constraint are, respectively,
\ba \tilde{{\cal{C}}}&:=&\frac{1}{2} \bigl(P_{\psi}^2 - 2\tau
P_{\tau}P_{\gamma}\bigr)+\frac{\tau}{2}\bigl(4 \tau''
-2\gamma'\tau'+\tau \psi'\,^2 \bigr)\,,\nonumber\\
\label{hamil-mom-const} {\cal{C}}_{\theta}&:=&P_{\tau}\tau'+
P_{\gamma}\gamma' + P_{\psi}\psi' - 2P_{\gamma}'\, . \ea The prime
denotes the derivative with respect to $\theta$.

Note that, since the spatial slices are compact, there exist no
boundary contributions to the Hamiltonian. Therefore, the total
Hamiltonian vanishes on the constraint surface and there is no
distinction between gauge and dynamics. It is then necessary to
carry out a deparametrization in order to introduce dynamics. This
deparametrization is accomplished as part of a gauge fixing of the
model: one imposes suitable conditions which, together with the
constraints (\ref{hamil-mom-const}), form a set of second class
constraints, allowing the reduction of the system. Explicitly, we
demand the following gauge-fixing conditions (which are a slight
modification of those introduced in Refs.
\cite{guillermo-gowdy,pierri}) \be \label{mom-cond}
g_{1}:=P_{\gamma}+p=0\,,\quad g_{2}:=\tau -tp=0 \, . \ee

The first of these conditions requires the momentum canonically
conjugate to $\gamma$ to be homogeneous (independent of $\theta$).
Furthermore, this homogenous part is a constant (of motion) $p$. In
this sense, it is worth pointing out that the Poisson brackets of
$\oint P_{\gamma}/(2\pi)$ (i.e. $-p$) with all the first class
constraints (\ref{hamil-mom-const}) vanish weakly, so that it is
indeed a Dirac observable. On the other hand, the second of our
conditions fixes the metric function $\tau$ equal to the global time
function $t$ except for a rescaling that is constant on shell,
though can vary on different solutions. Modulo constraints and
gauge-fixing conditions, a straightforward calculation shows then
that \be \label{2-class} \left\{g_{1},\oint d\theta \, G
{{\cal{C}}_{\theta}}\right\}=-pG' \,, \quad \left\{g_{2},\oint
d\theta \,\dfun \tilde{{\cal{C}}}\right\}=t p^2 \dfun  \, , \ee
where the smearing functions $\dfun$ and $G$ on $S^1$ are,
respectively, a density of weight $-1$ and a scalar. Therefore, if
$\dfun$ and $G'$ are different from zero, these Poisson brackets do
not vanish provided that $p\neq 0$. Thus, we have to restrict all
considerations to the sector of solutions with nonzero $p$ in order
to get a well-posed fixation.

The next step in this procedure consists in demanding the
compatibility of the gauge-fixing conditions with dynamics: the
total time derivative of $g_{1}$ and $g_{2}$ must vanish for some
choice of $N$ and $N^{\theta}$. This derivative is the sum of the
Poisson bracket with the total Hamiltonian $\oint \mathscr{H}$ and
the partial derivative with respect to the explicit $t$-dependence.
Modulo constraints and gauge-fixing conditions, we have \be
\dot{g}_{1}=-p\left(N^{\theta}\right)' \,, \quad
\dot{g}_{2}=-p+tp^2\dlapse \, . \ee The requirements $\dot{g}_{1}=0$
and $\dot{g}_{2}=0$ are then satisfied if $N^{\theta}$ is any
function of $t$ and $\dlapse = (tp)^{-1}$. It is worth noticing
that, while the densitized lapse function is completely determined
in this process, the shift function is not fully fixed. There
remains some diffeomorphism gauge freedom, generated by the
homogenous part of the constraint ${\cal{C}}_{\theta}$ (after
reduction). Besides, note that we have to further restrict $p$ to be
positive in order to ensure the positivity of the lapse
function\footnote{Otherwise, one should consider $t\in \R^{-}$
instead of $t\in \R^{+}$ for the time flow vector field to be future
directed.}.

In order to extract the true degrees of freedom, one solves the set
of second class constraints
$\{\tilde{{\cal{C}}},{\cal{C}}_{\theta},g_{1},g_{2}\}$, obtaining
\ba \label{tau-mom}
pP_{\tau}&=&-\frac{1}{2}\Biggl(\frac{P^{2}_{\psi}}{tp}+
tp\psi'\,^{2}\Biggr)\, , \\ \label{g-const-v1}
 p\gamma ' &=& P_{\psi}\psi' :=\Lambda\, . \ea
By performing a Fourier expansion in $\theta$ of the functions
$\gamma$ and $\Lambda$ (which is possible given the smoothness of
the fields on $S^1$), it is not difficult to see that identity
(\ref{g-const-v1}) allows us to solve for all modes of $\gamma$ but
the zero mode. More precisely, the Fourier coefficients $\gamma_{n}$
are determined in terms of $\Lambda_{n}$ by $inp\gamma_{n}=
\Lambda_{n}$. Thus, there is still an undetermined coefficient,
namely $\gamma_{0}$, and consequently we are left with a global
degree of freedom.

Furthermore, note that integration over $S^1$ of Eq.
(\ref{g-const-v1}) leads to the global constraint \be
\label{global-const} \Lambda_{0}=\frac{1}{\sqrt{2\pi}}\oint d\theta
\, P_{\psi} \psi '=0 \, , \ee which is essentially the homogenous
part of the constraint ${\cal{C}}_{\theta}$. Therefore, the
diffeomorphism gauge freedom has not been entirely removed and the
$\theta$-component of the shift vector cannot be completely fixed.
However, as we have already seen, the only allowed dependence of
$N^{\theta}$ is that on $t$. This type of shift can always be
absorbed by redefining our angular coordinate $\theta$
\cite{guillermo-gowdy}. After our gauge fixing and the absorption of
the shift, the metric becomes \ba \label{metric-po-gow}
ds^2&=&e^{\gamma - \psi}\left(-dt^2 + d\theta ^2\right) +
e^{-\psi}t^2p^2 d\sigma^2 + e^{\psi}d\delta^2\,,\\
\gamma&=&\frac{q}{2\pi}-i\sum_{n \neq
0}\frac{\Lambda_{n}}{np}\,\frac{e^{i n \theta}}{\sqrt{2\pi}}\,,\ea
where $q:=\sqrt{2\pi} \gamma_{0}$ is the coordinate canonically
conjugate to $-p$ (the zero mode of $P_{\gamma}/\sqrt{2\pi}$).

The reduced action for the system (modulo a spurious boundary term
$-pq \, \vert_{t_{i}}^{t_{f}}$) is
\begin{eqnarray}
\label{red-action-v1} S_{r}& = & \int_{t_{i}}^{t_{f}}dt\left\{
\dot{p}\left(q+\oint d \theta\, t P_{\tau}\right)+ \oint
d\theta\left[P_{\psi}\dot{\psi}+pP_{\tau}\right] \right\} \nonumber
\\ \, & = & \int_{t_{i}}^{t_{f}}dt \left\{ \dot{p} \eta + \oint d
\theta\left[P_{\psi}\dot{\psi}-\frac{1}{2}\left(\frac{P^{2}_{\psi}}
{tp}+tp\psi'\,^{2}\right) \right] \right\}, \nonumber\\
\eta&:=&q-\oint d \theta\, \frac{t}{2p}\left(\frac{P^{2}_{\psi}}
{tp}+tp\psi'\,^{2}\right).
\end{eqnarray}
In the first equality, $P_{\tau}$ denotes the solution given in Eq.
(\ref{tau-mom}). Thus, $S_{r}$ is a functional on the reduced phase
space $\G_{r}$, which is coordinatized by $(\eta , p , \psi ,
P_{\psi})$, and where the (only nonvanishing) basic Poisson brackets
are $\{p , \eta \}=1$ and
$\{\psi(t,\theta),P_{\psi}(t,\tilde{\theta})\}=\delta(\theta-
\tilde{\theta})$. Note that, owing to the presence of the global
constraint (\ref{global-const}), the space of physical states does
not correspond to $\G_{r}$ but to a submanifold of it. However,
since this submanifold is nonlinear, the reduction by the constraint
is usually postponed to the quantum theory, where it is imposed as
an operator condition on quantum states.

Let us now perform the canonical transformation \ba
\label{change-var} \quad \phi&=&\sqrt{p}\psi\,,\quad
P_{\phi}=\frac{P_{\psi}}{\sqrt{p}} \,,\nonumber\\
Q&=&-\eta+\frac{1}{2p}\oint d\theta \, P_{\psi}\psi  \,,\quad P=p\,
. \ea In terms of this new set of phase space variables the reduced
action reads \be \label{reduced-action}
S_{r}=\int_{t_{i}}^{t_{f}}dt\, \left( P\dot{Q} + \oint d
\theta\,\left[P_{\phi}\dot{\phi}-{{\hd}}_{r} \right] \right)
-PQ\big\vert_{t_{i}}^{t_{f}}\, , \ee where the (reduced) Hamiltonian
density is \be \label{reduced-hd} {{\hd}}_{r}=\frac{1}{2}
\left(\frac{P_{\phi}^{2}}{t}+t\phi'\, ^{2}\right) \, . \ee Thus, our
midisuperspace model consists of a phase space $\tilde{\G}_{r}$
coordinatized by the canonical pairs $(Q,P)$ and $(\phi ,
P_{\phi})$, which we will call the global and local degrees of
freedom, respectively. Remember that $P$ is strictly positive. To
arrive at a true canonical pair of real variables, we could always
replace $(Q,P)$ with $(QP,\ln{P})$. There also remains a global
constraint on the system ($\Lambda_0=0$) which restricts the
physical states to lie in a submanifold of $\tilde{\G}_{r}$. Note
that, given the $(Q,P)$-independence of the Hamiltonian density,
these ``point particle'' degrees of freedom are constants of motion.
Hence a nontrivial evolution may only take place in the field sector
$\G=\{(\phi , P_{\phi})\}$. Since the time evolution affects only
the local degrees of freedom, we will focus on them in our analysis.

Varying action (\ref{reduced-action}) with respect to $\phi$ and
$P_{\phi}$ one gets the field equations \be \label{canonical-f-eq}
P_{\phi}\,=\,t\,\dot{\phi} \, , \quad
\dot{P\,\,}\!_{\phi}\,=\,t\,\phi'' \, . \ee Hence, we only have to
consider all smooth solutions to the second-order differential
equation \be \label{eq-phi}
\ddot{\phi}+\frac{1}{t}\dot{\phi}-\phi''=0 \,  \ee in order to
specify the classical spacetime metric. Using the method of
separation of variables, it is not difficult to see that these
solutions, that we will generically denote by $\varphi$, adopt the
form \cite{torre-prd} \ba \label{scalarfield-sol}
\varphi(t,\theta)&=&\frac{1}{2\sqrt{2}}\sum_{n\in \Z,\,n\neq
0}\left[A_{n}H_{0}(|n|t)e^{in\theta}+A_{n}^{*}H_{0}^{*}(|n|t)
e^{-in\theta}\right] \nonumber \\ &+& \frac{1}{\sqrt{2\pi}}
\left(\bar{q}_{0}+\bar{p}_{0}\ln t\right)\, , \ea where the symbol
$*$ denotes complex conjugation, $\bar{q}_0$ and $\bar{p}_0$ are
constants, $H_{0}$ is the zeroth-order Hankel function of the second
kind \cite{abramowitz} and, in order to guarantee pointwise
convergence, the sequence of constant coefficients $\{A_{n}\}$ has
to decrease faster than the inverse of any polynomial in $n$ as
$n\to \pm \infty$. Expression (\ref{scalarfield-sol}) determines the
metric (\ref{metric-po-gow}) (with $\psi=\varphi/\sqrt{p}$) except
for the values of $(q,p)$. One can show that its Kretchmann scalar
blows up at $t=0$, so that there is an initial singularity and the
global time function $t$ must be strictly positive.

In the field sector, the physical phase space can be alternatively
described by the submanifold obtained by imposing the constraint
$\Lambda_{0}$ in $\G$, or by that submanifold of the space $V$ of
the smooth solutions (\ref{scalarfield-sol}) defined by the
constraint \be \label{global-const-cov}
\tilde{\,\Lambda}_{0}:=\sum_{n \in \Z, \,n\neq 0}n A^*_{n}A_{n}=0.
\ee In addition notice that, for the field sector, the reduced
action (\ref{reduced-action}) can be viewed as that corresponding to
an axi-symmetric, massless, free scalar field propagating in the
fictitious flat background in three dimensions: \be
\label{fictitious-metric}
^{(f)}g_{AB}=-(dt)_{A}(dt)_{B}+(d\theta)_{A}(d\theta)_{B}+t^{2}
(d\sigma)_{A}(d\sigma)_{B}\ . \ee Thus, we can identify $\G$ with
the canonical phase space of the scalar field in this background,
$({\cal M}\simeq T^{2}\times \R^+, ^{(f)}g_{AB})$, whereas the space
$V$ of smooth solutions can be considered as the covariant phase
space of such a Klein-Gordon field. Namely, $\phi$ and $P_{\phi}$
are the configuration and momenta on the constant-time section
$\Sigma_{t}$ of the scalar field $\varphi$ propagating in $({\cal
M},^{(f)}g_{AB})$. Besides, since the fictitious background is
globally hyperbolic, given a smooth Cauchy surface $\Sigma_{t_{0}}$
there will be a natural isomorphism $I[\Sigma_{t_{0}}]$ between the
linear spaces $\G$ and $V$. In this framework, the analysis of the
dynamics of the polarized Gowdy $T^{3}$ model becomes equivalent to
the study of the time evolution of the free scalar field. In the
next section we review this dynamics and discuss the obstructions to
its unitary quantum implementation.


\section{Dynamics}
\label{sec-dynamics}

It has recently been shown that the dynamical evolution generated by
the reduced Hamiltonian $H_{r}=\oint {{\hd}}_{r}$ [see Eq.
(\ref{reduced-hd})] cannot be implemented as a unitary
transformation, neither on the kinematical Fock space \cite{ccq-t3}
constructed from $V$ with the complex structure associated with the
field decomposition (\ref{scalarfield-sol}), nor in the physical
Hilbert space of states \cite{torre-prd} determined by the kernel of
the operator version of the constraint (\ref{global-const-cov}). We
want to analyze in detail the reasons behind this lack of a unitary
implementation and discuss how severe the problem is, studying
whether small corrections (coming e.g. from quantum or perturbative
modifications) to the dynamics may suffice to restore the unitarity.

 For the space $V$, we will employ as coordinates the constants
coefficients of the field decomposition (\ref{scalarfield-sol}),
whereas for $\G$ we will use a different set that absorbs in its
(implicit) time dependence all the evolution of the field. We will
see that the dynamics in $\G$ is dictated by $H_{r}$, whereas that
in $V$ is frozen, because the considered coefficients are constants
of motion {\footnote{Let us emphasize that the coefficients in Eq.
(\ref{scalarfield-sol}) do not display any time dependence, not only
explicitly, but also implicitly. Accordingly, the total Hamiltonian
in $V$ indeed vanishes. Properly speaking, the nonunitarity proved
in Ref. \cite{ccq-t3} is that of the transformation generated by
$H_{r}$ on $V$, which in turns can be seen to imply the nonunitary
character of the dynamics in $\G$, rather than in $V$.}}.

 For the sake of completeness and clarity, let us remember some
definitions and make a few remarks that will be useful in our
analysis. Firstly, we recall that given two field decompositions in
different orthonormal bases of solutions, namely
$\varphi=\sum_{n}\!A_{n}f_{n}(t,\theta)+A_{n}^{*}f_{n}^{*}
(t,\theta)$ and
$\varphi=\sum_{n}\!\tilde{A}_{n}g_{n}(t,\theta)+\tilde{A}_{n}^{*}
g_{n}^{*}(t,\theta)$, their coefficients are related by a Bogoliubov
transformation. That is,
$\tilde{A}_{n}=\sum_{m}\alpha_{mn}A_{m}+\beta_{mn}^{*}A_{m}^{*}$
with $\sum_{k}(\alpha_{ik}\alpha_{jk}^{*}-\beta_{ik}\beta_{jk}^{*})=
\delta_{ij}$ and
$\sum_{k}(\alpha_{ik}\beta_{jk}-\beta_{ik}\alpha_{jk})=0$
\cite{birrel-davies}.

Secondly, as in the classical case, the quantum theory for the
scalar field can be formulated either in a covariant or in a
canonical approach. In fact, it is generally known that, by endowing
the space of (smooth) solutions $V=\{\varphi\}$ to the Klein-Gordon
equation with a complex structure $J$ compatible with the symplectic
structure, one can construct in a canonical way the Hilbert space of
the quantum theory as a symmetric Fock space ${\fock}$ on which the
basic observables of the theory are represented as annihilation and
creation operators \cite{wald}. On the other hand, if the scalar
field propagates in a globally hyperbolic spacetime, e.g. ${\cal
M}\approx \Sigma \times \R^+$, given an embedding $\Sigma_{t_{0}}$
of $\Sigma$ as a Cauchy surface in ${\cal M}$ one gets, from the
covariant complex structure $J$ on $V$, the induced complex
structure $J_{0}$ on the canonical phase space $\G$ \cite{ccq-prd}.
Once we have $J_{0}$, we know how to construct the Schr\"{o}dinger
representation which is unitarily equivalent to the Fock one
\cite{ccq-prd} and how to pass from one representation to the other
\cite{ccq-annals}. In particular, we have the analog of the
one-particle Hilbert space and, by applying the creation operator,
one can construct the $n$ (functional) particle states.

In addition, as a consequence of this unitary relation between the
covariant and canonical approaches, if a symplectic transformation
is unitarily implementable with respect to the Fock representation,
so is it with respect to the (unitarily equivalent) Schr\"{o}dinger
representation (and vice versa). Recall that a symplectic
transformation on $V$ ($\G$) will be unitarily implementable with
respect to the Fock (Schr\"{o}dinger) representation if the
antilinear part of its quantum counterpart is Hilbert-Schmidt on the
one-particle Hilbert space \cite{honegger}. For a quantum
transformation of the Bogoliubov type, the Hilbert-Schmidt condition
reduces to $\sum_{n,m}| \beta_{nm}|^{2}< \infty $.

Consider now the Schr\"{o}dinger representation constructed from
$J_{0}$ (the complex structure induced by the covariant one, $J$).
Given a symplectic transformation $T$ on $\G$, we obtain an induced
complex structure $J_{0}'=T J_{0} T^{-1}$ and, associated with it, a
new Schr\"{o}dinger representation. The annihilation and creation
operators in this new representation are related with the
annihilation and creation operators of the former one through a
Bogoliubov transformation. The representations corresponding to
$J_{0}'$ and $J_{0}$ are then unitarily equivalent if the antilinear
part of this Bogoliubov transformation is square summable.

In particular, if $T=T_{(t_{1},t_{2})}$ represents the time
evolution from $t_{1}$ to $t_{2}$ on $\G$, then the Bogoliubov
transformation relates the annihilation  and creation operators at
the instants $t_{1}$ and $t_{2}$. With an appropriate choice of
coordinates in $\G$, the symplectic transformation
$T_{(t_{1},t_{2})}$ acts on the elements of this space exactly as
the Bogoliubov transformation on their quantum counterparts. Thus,
in order to elucidate whether the symplectic transformation is
unitarily implementable it suffices to analyze the square
summability of the antilinear part of $T_{(t_{1},t_{2})}$. Let us
emphasize that this procedure is equivalent to determining whether
the Schr\"{o}dinger representations constructed from the same $\G$
(associated with the embedding $\Sigma_{t_{1}}$) but with the
distinct complex structures $J_{1}$ and $J_{2}'=T_{(t_{1},t_{2})}
J_{1} T_{(t_{1},t_{2})}^{-1}$ are unitarily related.

In the following, we will focus our discussion on analyzing the
failure of unitary implementability of the symplectic transformation
that determines the dynamics in $\G$, rather than examining the
complex structures induced by it, since both procedures are
equivalent, as we have just seen.

\subsection{Evolution in the canonical and covariant approaches}
\label{sec-coordinates-approaches}

We start by expanding in Fourier series our canonical variables
$\phi$ and $P_{\phi}$: \be \label{fourier-series} \phi=\sum_{n\in
\Z}\frac{1}{\sqrt{2\pi}}\phi_{n}e^{in\theta} \, , \quad
P_{\phi}=\sum_{n\in \Z}\frac{1}{\sqrt{2\pi}} P_{\phi}^{-n}
e^{in\theta} \, , \ee where the coefficients $\phi_{n}$ and
$P_{\phi}^{n}$ are (implicit) functions of the global time
coordinate. From the basic Poisson bracket between $\phi$ and
$P_{\phi}$, it is easily shown that
$\{\phi_{n},P_{\phi}^{m}\}=\delta_{n}^{m}$. Therefore, we can
equivalently consider as our canonical phase space that whose
coordinates are the set of (complex) canonical pairs
$\{(\phi_{n},P_{\phi}^{n})\}_{n\in \Z}$. We will call it
${\G}_{(\phi,P_{\phi})_{n}}$. Note that this space can be decomposed
as the direct sum of ${\G}'$ and ${\G}_{0}$, where ${\G}'$ is the
subspace of vectors with $\phi_{0}=P_{\phi}^{0}=0$ and ${\G}_{0}$ is
the span of those vectors whose only nonvanishing components are
precisely those corresponding to $\phi_{0}$ and $P_{\phi}^{0}$. For
all positive integers $m\in \N -\{0\}$, let us now consider the
transformations \be \label{cano-trans-gamma} (\phi_{m} ,
P_{\phi}^{m} , \phi_{-m} , P_{\phi}^{-m})\mapsto (a_{m} , a_{m}^{*}
, a_{-m} , a_{-m}^{*})\, , \ee where \be \label{anh-cre-canonical}
a_{m}=\frac{| m | \phi_{m}+iP_{\phi}^{-m}}{\sqrt{2| m|}}\,, \quad
a_{m}^{*}=\frac{| m | \phi_{-m}-iP_{\phi}^{m}}{\sqrt{2| m|}}\,. \ee

One can check that these transformations are canonical, so that
$\{a_{n},ia_{m}^{*}\}=\delta_{nm}$. Hence, the canonical phase
space can be alternatively described by the symplectic vector
space ${\G}_{a}={\G}_{0}\oplus \bar{\G}$, where the coordinates
for $\bar{\G}$ are the (complex conjugate pairs of) annihilation
and creation-like variables $\{(a_{m} ,a_{m}^{*}, a_{-m} ,
a_{-m}^{*})\}_{m\in \N -\{0\}}$. The dynamics on $\G_{a}$ [as well
as in $\G$ and ${\G}_{(\phi,P_{\phi})_{n}}$] is dictated by the
reduced Hamiltonian for the polarized Gowdy $T^{3}$ model,
$H_{r}=\oint {{\hd}}_{r}$, where ${{\hd}}_{r}$ is the reduced
Hamiltonian density (\ref{reduced-hd}). Using expressions
(\ref{fourier-series}) and (\ref{anh-cre-canonical}), and defining
\ba \label{our-hamil}
H_{0}&:=&\frac{\bigl(P_{\phi}^{0}\bigr)^{2}}{2t} \,, \nonumber\\
H_{m}&:=&\frac{t^{2}+1}{2t}\,m\,(a_{m}a_{m}^{*}+a_{-m}a_{-m}^{*})
\nonumber\\
&+&\frac{t^{2}-1}{2t}\,m\,(a_{m}^{*}a_{-m}^{*}+a_{m}a_{-m}) \,, \ea
the Hamiltonian can be rewritten \be \label{hamiltonian-gamma}
H_{r}= H_0[P_{\phi}^0]+ \!\!\sum_{m\in \N,\,m\neq
0}\,H_{m}[a_m,a^{*}_m,a_{-m},a^{*}_{-m}].\ee

Notice that the Hamiltonian vector field $X^{A}_{H_{r}}$ on
${\G}_{a}$ is just the sum of the Hamiltonian vector fields
$X^{A}_{H_{0}}$ on ${\G}_{0}$ and $X^{A}_{H}$ on $\bar{\G}$, with
$H:=\sum_{m\in\N-\{0\}}H_m$. In other words, for any state in
${\G}_{a}$, the time evolution can be deduced by composing the
evolution of the projections in ${\G}_{0}$ and in $\bar{\G}$, where
the dynamics are dictated, respectively, by $H_{0}$ and $H$. In
particular, we see that the unitary implementability of the dynamics
in ${\G}_{a}$ at the quantum level depends only on whether the
finite transformations generated by $H$ on $\bar{\G}$ can be
unitarily implemented.

The covariant phase space $V$, on the other hand, can be described
using as coordinates the canonical pair $(\bar{q}_0,\bar{p}_0)$ and
the set of (complex conjugate) annihilation and creation-like
variables $\{(A_{n},A_{n}^{*})\}_{n\in \Z-\{0\}}$. Let us denote the
covariant phase space, in such a coordinate system, by $V_{A}$. We
can now separate the zero modes exactly as before, namely, $V_{A}$
can be viewed as the direct sum of the subspace $\bar{V}$ for which
$\bar{q}_0=\bar{p}_0=0$ and the subspace $V_{0}$ whose vectors have
$\bar{q}_0$ and $\bar{p}_0$ as the only nonzero components.

In the following, we will respectively denote states in ${\G}_{0}$
and $V_{0}$ by $u_{0}:=(\phi_{0},P_{\phi}^{0})$ and
$v_{0}:=(\bar{q}_0,\bar{p}_0)$. Similarly, states in $\bar{\G}$ and
$\bar{V}$ will be denoted by $\{\gamma_{m}\}:=\{(a_{m} ,a_{m}^{*} ,
a_{-m} , a_{-m}^{*})\}$ and $\{{\cal{A}}_{m}\}:=\{(A_{m} , A_{m}^{*}
, A_{-m} , A_{-m}^{*})\}$, with $m\in \N-\{0\}$.

Since $\phi$ and $P_{\phi}$ are the configuration and momenta at
$\Sigma_{t}$ of $\varphi$, we can express the coefficients
$\phi_{n}$ and $P_{\phi}^{-n}$ in terms of $A_{n}$ and $A_{-n}^{*}$
for all $n\in \Z-\{0\}$, and in terms of $\bar{q}_0$ and $\bar{p}_0$
for the zero mode, getting in this way a map $\tilde{M}$ from
$V_{A}$ to ${\G}_{(\phi,P_{\phi})_{n}}$. In addition, Eq.
(\ref{anh-cre-canonical}) defines a map $\,\overline{\!M}$ from
${\G}_{(\phi,P_{\phi})_{n}}$ to ${\G}_{a}$. Then the composition
$\,\overline{\!M}\tilde{M}$ provides us with a map $M:V_{A} \to
{\G}_{a}$. A straightforward calculation shows that for the zero
modes
\begin{eqnarray}
\label{zero-modes-map} u_{0}=\left(
\begin{array}{cc} 1\, & \, \ln t  \\ 0 & 1\end{array}
\right) v_{0} \, , \end{eqnarray} where $u_0$ and $v_0$ are treated
as column vectors. For the rest of modes ($m\in \N-\{0\}$), one
obtains $\gamma_{m}=M_{m}(t)\,{\cal{A}}_{m}$ with \be
\label{nonzer-modes-map} M_{m}(t)= \left(
\begin{array}{cccc} c_{m}(t) & 0 & 0 & d_{m}(t) \\ 0
& c_{m}^{*}(t) & d_{m}^{*}(t) & 0 \\ 0 & d_{m}(t) & c_{m}(t) & 0
\\ d_{m}^{*}(t) & 0 & 0 & c_{m}^{*}(t) \end{array} \right)  \ee
and \ba c_{m}(t)&=&\sqrt{\frac{\pi m}{8}}
\left[H_{0}(mt)-itH_{1}(mt)\right],\nonumber\\
d_{m}(t)&=&\sqrt{\frac{\pi m}{8}}\left[H_{0}^{*}(mt)
-itH^{*}_{1}(mt)\right].\ea Here, $H_{1}$ is the first-order Hankel
function of the second kind \cite{abramowitz}. Note that the map $M$
is such that $M(V_{0})={\G}_{0}$ and $M(\bar{V})={\bar{\G}}$.
Besides, the determinant of the linear transformation
(\ref{zero-modes-map}), as well as that of $M_{m}$, is equal to the
unity. It hence follows that the $M_{m}$'s are Bogoliubov
transformations. Thus, we get a time-dependent canonical
transformation from $V_{A}$ to ${\G}_{a}$.

A generating function for this transformation (that depends on some
appropriately chosen complete sets of compatible components -under
Poisson brackets- both for $V_A$ and $\G_a$) is \ba
\label{generating-func} {\cal F}_{0}(t)&=&
\frac{1}{2}(\bar{p}_0)^{2}\ln t - \bar{p}_{0}\phi_{0}, \nonumber
\\ {\cal{F}}_{m}(t)&=&
ia_{-m}^{*}\bigl[c_{m}(t)A_{-m}+d_{m}(t)A_{m}^{*}\bigr]\nonumber\\
&-&ia_{m}\bigl[ d^{*}_{m}(t)A_{-m}+c^{*}_{m}(t)A_{m}^{*}\bigr]\,,
\ea for $m=0$ and $m \in \N -\{0\}$, respectively. After a
straightforward calculation we find that the partial derivative of
this generating function with respect to its explicit dependence
on the time coordinate $t$ has the following form when expressed
exclusively in terms of the components of the states in
${\G}_{a}$: \be \label{deriv-generating-func}
\partial_{t}{\cal{F}}_{0}=
\frac{1}{2t}\bigl(P_{\phi}^{0}\bigr)^{2} \, , \quad
\partial_{t}{\cal{F}}_{m}=H_{m}[\gamma_m] \, . \ee
Therefore, we get \be \label{gene-hd}
\partial_{t}{\cal{F}}[u,\gamma]=\sum_{m\in \N}\partial_{t}
{\cal{F}}_{m}[u,\gamma] \, = \, H_{r}[u,\gamma] \, , \ee where
$H_{r}$ is precisely the Hamiltonian (\ref{hamiltonian-gamma}).

At this point of the discussion, it is worth recalling that, given a
canonical transformation from certain symplectic vector space
$E_{1}:=\{(q_{i},p_{i})\}$ to another one $E_{2}:=\{(Q_{i},P_{i})\}$
which is determined by a generating function $F$ that is explicitly
time dependent \cite{goldstein}, and assumed that the dynamics in
$E_1$ is dictated by the Hamiltonian $H_1[q,p]$, the corresponding
Hamiltonian in $E_2$ is
$H_{2}[Q,P]=H_{1}[q(Q,P),p(Q,P)]+\partial_{t}F[Q,P]$. Taking into
account Eq. (\ref{gene-hd}), we then see that the dynamical
evolution in $\G_a$, generated by $H_r$, arises entirely from the
time dependence of the canonical transformation. As we have pointed
out before, the total Hamiltonian in $V_{A}$ is identically zero and
there is no time evolution for the states
$(v_{0},\{{\cal{A}}_{m}\})$. Obviously, this vanishing of the
Hamiltonian applies as well to the restrictions to the subspaces
$V_{0}$ and $\bar{V}$. In particular, while the states in $\bar{\G}$
evolve along the integral curves of the Hamiltonian vector field
$X^{A}_{H}$, the states in $\bar{V}$ are ``frozen''. Hence, an
initial state $\{\gamma_{m}(t_{0})\}$ in $\bar{\G}$ will evolve to
the state $\{\gamma_{m}(t)\}$ determined by the transformation
$\gamma_{m}(t)=U_{H_{m}[\gamma]}^{(m)}(t,t_{0})\gamma_{m}(t_{0})$,
where\footnote{Actually
$U_{H_{m}[\gamma]}^{(m)}(t_{0},t_{0})={\rm{I}}$ and
$U_{H_{m}[\gamma]}^{(m)}(t_{3},t_{1})=U_{H_{m}[\gamma]}^{(m)}(t_{3},
t_{2})U_{H_{m}[\gamma]}^{(m)}(t_{2},t_{1})$.}
$U_{H_{m}[\gamma]}^{(m)}(t,t_{0})\!:=M_{m}(t)M_{m}(t_{0})^{-1}$. In
contrast, the corresponding states in $\bar{V}$, specified by
${\cal{A}}_{m}(t_{0})=M_{m}(t_{0})^{-1}\gamma_{m}(t_{0})$ and
${\cal{A}}_{m}(t)=M_{m}(t)^{-1}\gamma_{m}(t)$, will be related via
the identity map, so that they actually coincide.

In coordinates ${\cal{A}}_{m}$ rather than $\gamma_m$, the finite
transformation $U_{H_{m}[\gamma]}^{(m)}$ is given by
$U_{H_{m}[{\cal{A}}]}^{(m)}(t,t_{0})=M_{m}(t_{0})^{-1}M_{m}(t)$. As
a result, the complex antilinear part of the finite transformation
generated by $H=H_r-H_0$ in $\bar{V}$ is given (for each $m\in
\N-\{0\}$) by \be \label{antili-in-v} D_{m}(t,t_{0})=\frac{i\pi
m}{4}\left[t_{0}H_{1}^{*}(mt_{0})H_{0}^{*}(mt)-tH_{0}^{*}(mt_{0})
H_{1}^{*}(mt)\right] , \ee which is not square summable in $m$, as
has been proved in Ref. \cite{ccq-t3}. Therefore, the finite
transformation provided by $M_{m}(t_{0})^{-1}M_{m}(t)$ (with $m$
running in $\N-\{0\}$) cannot be unitarily implemented. Moreover,
since the antilinear part of $U_{H_{m}[\gamma]}^{(m)}(t,t_{0})$
differs from that of $U_{H_{m}[{\cal{A}}]}^{(m)}(t,t_{0})$ just by a
sign in the phase of the coefficient $c_{m}(t)$, as one can easily
check, we see that the finite transformation generated by $H$ in
$\bar{\G}$ is not unitarily implementable. It is worth emphasizing
that, however, this it is not the case for the dynamics in
$\bar{V}$; indeed, since such a dynamics is generated by the zero
Hamiltonian, the evolution is described by the identity
transformation, which is of course unitary.

Actually, since $U_{H_{m}[\gamma]}^{(m)}(t,t_{0})$ is just a
composition of the Bogoliubov transformations $M_m(t)$ and
$M_m(t_0)$, the lack of a unitary implementation of the dynamics in
$\bar{\G}$ follows from the fact that the antilinear part of
$M_{m}(t)$ fails to be square summable for generic $t>0$. Hence,
whether or not the dynamics can be unitarily implemented depends
entirely on the behavior of $d_{m}(t)$ for large integers $m$. This
depends in turn on the Hamiltonian via Eq.
(\ref{deriv-generating-func}), which relates the generating function
of $M_{m}(t)$ with the generator of the dynamics after the canonical
transformation has been performed. Our analysis about the lack of
unitarity will therefore focus on the identification of those
characteristics of the Hamiltonian that are in the origin of the
failure of square summability. In doing so, we will be able to
establish how critical this problem is and whether it can be
corrected with small modifications to the dynamics. Roughly
speaking, we will be able to determine how far the considered
evolution is from being unitarily implementable. With this aim, in
the next subsection we introduce a correction to the Hamiltonian
$H$, which might be viewed as a perturbation or a quantum
modification, and discuss the square summability of the antilinear
part of the transformation generated by the new Hamiltonian.

\subsection{Modified Hamiltonian}
\label{sec-pert-ham}

Motivated by our previous analysis, let us assume now that a certain
linear (free) field theory can be described by either of the two
symplectic vector spaces $E_{A}:=\{{\cal A}_{m}\}$ or
$E_{B}:=\{{\cal B}_{m}\}$, where ${\cal
A}_{m}:=(A_{m},A_{m}^{*},A_{-m},A_{-m}^{*})$, ${\cal
B}_{m}:=(B_{m},B_{m}^{*},B_{-m},B_{-m}^{*})$, and $m\in \N-\{0\}$.
Here, $(A_{m},A_{m}^{*})$ and $(B_{m},B_{m}^{*})$ are annihilation
and creation-like pairs. In addition, let us suppose that the
canonical map $M_m(t)$ from $E_{A}$ to $E_{B}$, which in general may
depend on the time coordinate $t$, is a Bogoliubov transformation of
the form (\ref{nonzer-modes-map}), with $|c_{m}(t)|^{2} -
|d_{m}(t)|^{2}=1$. Besides, we assume that the total Hamiltonian in
$E_{A}$ is zero, so that the Hamiltonian in $E_{B}$ is given by
$\tilde{H}=\sum_{m\in \N -\{0\}}\partial_{t}{\cal{F}}_{m}$, where
${\cal{F}}_{m}$ is a generating function of the transformation
$M_m(t)$. Furthermore, we admit that (in coordinates ${\cal A}_{m}$)
the partial derivative of the generating function ${\cal{F}}_{m}$
with respect to its explicit time dependence is \ba
\label{gene-der-imp}
\partial_{t}{\cal F}_{m}= \tilde{H}_{m}[{\cal A}]:&=&2m \left[\mu
a(x)A_{m}A_{-m}+\mu^{*}
a^{*}(x)A_{m}^{*}A_{-m}^{*}\right]\nonumber\\
&+&2m \lambda b(x)\left[A_{m}^{*}A_{m}+A_{-m}^{*}A_{-m}\right], \ea
where $x:=mt$ (strictly speaking we should write $x_{m}$, however we
drop out the subindex to simplify the notation). Besides, $\lambda$
and $\mu$ are a real and a complex constant\footnote{We might allow
for an $x$-dependence in $\lambda$ and $\mu$, but this would
unnecessarily complicate our discussion. Some comments about this
generalization of the analysis are presented in Subsec. III.C.},
respectively, and \ba \label{abform}
a(x)&=&\frac{\pi}{8}x\left\{\left[H_{0}(x)\right]^{2}+
\left[H_{1}(x)\right]^{2}\right\}, \nonumber
\\
b(x)&=&\frac{\pi}{8}x\left\{\left| H_{0}(x)\right|^{2}+ \left|
H_{1}(x)\right|^{2}\right\}. \ea

At this stage, it is convenient to point out the analogy with our
symmetry-reduced model. The Bogoliubov transformation $M_m(t)$ has
the same form as that in Eq. (\ref{nonzer-modes-map}). The
symplectic vector spaces $E_{A}$ and $E_{B}$ play the role of
$\bar{V}$ and $\bar{\G}$, respectively. In this sense, note that the
total Hamiltonian in $E_{A}$ is zero. Moreover, setting
$\lambda=\mu=1$ in equation (\ref{gene-der-imp}) one merely gets the
$H_{m}$ contribution to the total Hamiltonian $H$ in $\bar{\G}$,
expressed in coordinates ${\cal{A}}_{m}$ \cite{torre-prd}. Hence, we
can think in terms of the spaces $\bar{V}$ and $\bar{\G}$, and
regard the phase space function $\tilde{H}$ as a modification of the
Hamiltonian $H$. Defining $\rho:=\lambda-1$ and $\epsilon=\mu-1$,
one may view the case $|\rho|\ll 1$, $|\epsilon|\ll 1$ as a
perturbation of the Hamiltonian, arising from certain (classical or
quantum) corrections to the dynamics.

As we already know, the unitary implementability of the dynamics
dictated by $\tilde{H}$ in $E_{B}$ depends on the square summability
of the antilinear part of $M_{m}(t)$. By analyzing this summability
we will relate the failure of unitarity with the precise form of
$\tilde{H}$.

We first determine the relations that Eq. (\ref{gene-der-imp})
imposes on the complex functions $c_{m}(t)$ and $d_{m}(t)$ that
specify the Bogoliubov transformation $M_m(t)$. A generating
function for this transformation is \ba \label{gene-func-general}
{\cal{F}}_{m}(t)&=&iB_{-m}^{*}\left[c_{m}(t)A_{-m}+d_{m}(t)A_{m}^{*}
\right]\nonumber\\
&-&iB_{m}\left[d^{*}_{m}(t)A_{-m}+c^{*}_{m}(t)A_{m}^{*} \right]. \ea
Taking the (explicit) time derivative and using then the inverse of
$M_{m}(t)$ [which is easily calculated from Eq.
(\ref{nonzer-modes-map})], one arrives at the following expression,
exclusively in terms of the coordinates ${\cal B}_m$,
\begin{eqnarray}
\label{gene-func-der-general}
\!\!\!\!\partial_{t}{\cal{F}}_{m}\!&=&\!iB^{*}_{-m}
B_{-m}(\dot{c}_{m}c_{m}^{*}-\dot{d}_{m}d_{m}^{*})
+iB_{m}B_{-m}(\dot{c}_{m}^{*}d_{m}^{*}-\dot{d}_{m}^{*}c_{m}^{*})
\nonumber \\ \!\!\!\!& + & \!iB^{*}_{m}B_{m}
(\dot{d}_{m}^{*}d_{m}-\dot{c}_{m}^{*}c_{m})+iB^{*}_{m}B^{*}_{-m}
(\dot{d}_{m}c_{m}-\dot{c}_{m}d_{m}).
\end{eqnarray}
The dot denotes the (total) derivative with respect to the time
coordinate $t$. From now on, we do not generally display the
dependence of $c_m$ and $d_m$ on this coordinate in order to
simplify the notation.

By translating also into coordinates ${\cal B}_m$ the Hamiltonian
$\tilde{H}_m$, the condition (\ref{gene-der-imp}) that the dynamics
in $E_{B}$ arise entirely from the time derivative of our canonical
transformation can be seen to reduce to the following system of
first-order (complex) differential equations for $c_{m}$ and
$d_{m}$: \ba
0&=&\frac{i}{2}\left(\dot{c}_{m}^{*}c_{m}-\dot{d}_{m}^{*}d_{m}
\right)
+m\,\lambda b(x)\left(|c_{m}|^{2}+|d_{m}|^{2}\right)\nonumber
\\&-&m\left[\mu a(x)c_{m}^{*}d_{m}+\mu^*a^{*}(x)c_{m}
d_{m}^{*}\right] , \label{cond1} \\
0&=&\frac{i}{2}\left(\dot{d}_{m}c_{m}-\dot{c}_{m}
d_{m}\right)+2m\lambda b(x)
c_{m}d_{m}\nonumber\\
&-&m\left[\mu a(x)d_{m}d_{m}+\mu^* a^{*}(x)c_{m} c_{m}\right] .
\label{cond2}\ea

Let us call $Y_{m}$ the ratio $d_{m}/c_{m}$. Since $|c_{m}|^{2}
-|d_{m}|^{2}=1$, we have that $|Y_{m}|$ is strictly smaller than the
unity and, in terms of it, the (complex) norms of $c_{m}$ and
$d_{m}$ are \be \label{c-d-y} |c_{m}|^{2}=\frac{1}{1-|Y_{m}|^{2}} \,
, \quad |d_{m}|^2=\frac{|Y_{m}|^{2}}{1-|Y_{m}|^{2}} \, . \ee
Realizing that $(\dot{d}_{m}c_{m}-\dot{c}_{m}d_{m})/{c}_{m}^{2}$ is
just the time derivative of $Y_m$ and performing the change of
variable $t \mapsto x=mt$ (so that $\dot{Y}_{m}=m\,dY_{m}/dx$), it
is easy to see that Eq. (\ref{cond2}) can be rewritten as \be
\label{eq-y} \frac{i}{2}\frac{dY_{m}}{dx}(x)+2\lambda b(x)
Y_{m}(x)-\mu a(x)Y_{m}^{2}(x)-\mu^*a^{*}(x)=0\,. \ee Remarkably,
this differential equation for $Y_m$ is independent of the positive
integer $m$ (regarding $x$ as the relevant variable). Using this
universal character of the equation, valid for all values of $m$, we
can drop out the subindex in the function $Y_m$ and consider it as a
single function $Y$ for all the modes of our system. With the
convenient redefinitions \be \label{z} z(x):=\exp{(-2ix)}\,
Y(x)\,,\quad \Delta(x):=\exp(2ix)\, a(x)\,,\ee we then arrive at the
following equation for $z$: \be \label{exact-z-eq} \frac{dz}{dx}(x)
= 2i z(x)\left[ 2\lambda b(x)-1\right]-2i\mu\Delta(x)
z^{2}(x)-2i\mu^*\Delta^{*}(x)\,. \ee

In addition, given a function $z$ satisfying Eq. (\ref{exact-z-eq})
and remembering that $|c_m|^2-|d_m|^2=1$, it is straightforward to
see that the differential equation (\ref{cond1}) is equivalent to
\be \label{eqc}
\frac{d\,\ln{c_{m}}}{dx}(x)=2i\left[\mu\Delta(x)z(x)-\lambda
b(x)\right]\ee which again is a universal equation for all modes
$m\in\N-\{0\}$. We suppress the subindex $m$ and consider only one
function $c(x)$, which can be obtained by direct integration of Eq.
(\ref{eqc}) [except for a multiplicative constant that can be fixed
with an initial condition for $c$]. Finally, the function
$d(x):=\exp{(2ix)}z(x)c(x)$ provides the missing coefficient of our
Bogoliubov transformation, namely, $d_{m}(t)=d(x=mt)$.

Using Eq. (\ref{c-d-y}) and $|Y(x)|=|z(x)|$, we conclude
\be\label{sqdm}\left|d_m(t)\right|^2
=\left|d(x=mt)\right|^2=\frac{\left|z(x=mt)\right|^2}{1-\left|z(x=mt)
\right|^2}.\ee Therefore, an important consequence of the observed
universality is that the square summability of the coefficients
$d_m(t)$ at any fixed positive value of $t$, which is only sensitive
to the behavior for large $m$, turns out to depend exclusively on
the behavior of the function $z(x)$ when $x$ approaches infinity
(because $x=mt$ grows linearly with $m$ for all $t>0$). Thus, to
discuss the square summability of $d_m$, we only need to consider
Eq. (\ref{exact-z-eq}) and exploit our knowledge about the
asymptotic behavior of the functions $\Delta(x)$ and $b(x)$.

From Hankel's asymptotic expansions of $H_{0}$ and $H_{1}$
\cite{abramowitz} one gets that, for $x\gg 1$, $\Delta(x)$ and
$b(x)$ are given by the asymptotic series{\footnote{In general, we
will say that a function $f$ admits an asymptotic series at infinity
if there exists a series of the form
$\sum_{k=0}^{\infty}f_{k}/x^{k}$ such that $\lim_{x \to
\infty}\left|
x^{N}\left\{f(x)-\sum_{k=0}^{N}f_{k}/x^{k}\right\}\right|=0$ for all
$N\geq 0$ (see e.g. Ref. \cite{bender-orszag}).}}
\begin{eqnarray}
\label{asympt-b-delta-1}  \Delta(x) \!& = &
\!\frac{i}{4}\sum_{k,n=0}^{\infty} \frac{(-1)^{k+n}}{(2x)^{2(k+n)}}
\left[\xi_{k,n}
-i\frac{\xi_{k,n+1/2}}{x}-\frac{\xi_{k+1/2,n+1/2}}{4x^{2}}\right],
\nonumber\\
b(x)\! & = &\! \frac{1}{4}\sum_{k,n=0}^{\infty}
\frac{(-1)^{k+n}}{(2x)^{2(k+n)}}
\left[\sigma_{k,n}+\frac{\sigma_{k+1/2,n+1/2}} {4x^{2}}\right]\,
\end{eqnarray}
where \ba \xi_{k,n}&:=&(0,2k)(0,2n)-(1,2k)(1,2n)\,,\nonumber\\
\sigma_{k,n}&:=&(0,2k)(0,2n)+(1,2k)(1,2n)\,, \ea and $(k,n)$ is the
so-called Hankel symbol: \be
(k,n):=\frac{\G\left(k+n+\frac{\,1}{\,2}\right)}{n! \, \G\left(k -
n+\frac{\,1}{\,2}\right)}.\ee

The asymptotic series representation for $b(x)$ contains only even
powers in $1/x$, while the corresponding asymptotic series for
$\Delta$ contains, in principle, both even and odd powers. With the
change of variable $y=1/x$, these asymptotic expansions can thus be
written in the form \be \label{asympt-b-delta-2} \Delta(y) =
\sum_{k=0}^{\infty}\Delta_{k}y^{k}\, , \quad b(y) =
\sum_{k=0}^{\infty}b_{k}y^{k}\,, \ee where $b_{2k+1}=0$ for all
$k\in \N$. From Eqs. (\ref{asympt-b-delta-1}), we get in particular
\ba \label{coefbD} \Delta_{0}&=&i\,\frac{\xi_{0,0}}{4}=0\,,\quad
\Delta_{1}=\frac{\xi_{0,1/2}}{4}=-\frac{\,1}{\,4}\,,
\nonumber\\
b_{0}&=&\frac{\sigma_{0,0}}{4}=\frac{\,1}{\,2}\,.\ea

Employing the asymptotic expressions for $\Delta(x)$ and $b(x)$, a
formal asymptotic series for $z$ can be constructed. More precisely,
introducing expansions (\ref{asympt-b-delta-2}) in the differential
equation (\ref{exact-z-eq}) (with the change $y=1/x$) and writing
$z$ as an asymptotic series in $y$, namely $z=\sum_k z_k y^k$, one
obtains \be \label{eq-series}
z^{2}_{0}\mu\Delta_{0}+z_{0}(1-2\lambda b_{0})
+\mu^*\Delta^{*}_{0}+\sum_{k=1}^{\infty} y^{k} S_k=0\,,\ee where,
for each $k\geq 1$, \ba\label{S-series} S_k&:=&
\mu\sum_{m=0}^{k}\Delta_{k-m}
\biggl(\sum_{j=0}^{m}z_{m-j}z_{j}\biggr)-2\lambda
\sum_{m=0}^{k}b_{m}z_{k-m}  \nonumber\\
&+&z_{k}+\mu^*\Delta^{*}_{k}
+i\,\frac{k-1}{2}z_{k-1}+\mu\Delta_{k}z^{2}_{0}\,. \ea In
particular, the term independent of $y$ in Eq. (\ref{eq-series})
must vanish. Since $b_{0}=1/2$ and $\Delta_{0}=0$, one gets (with
$\rho=\lambda-1$) \be \label{z-o-coeff} z_{0}\, \rho =0 \, . \ee
Hence, provided that $\rho$ does not vanish, the coefficient $z_{0}$
must be zero. As a consequence, the resulting function $z$ will tend
to zero as $x \to \infty$, which in turn implies that $|d|$ vanishes
in the limit of large values of $x$.

The rest of terms in Eq. (\ref{eq-series}) require the vanishing of
$S_k$ for all $k\geq 1$. Substituting $z_{0}=0$ and the values of
$b_{0}$ and $\Delta_{0}$, it is straightforward to derive the
following recurrence relation for the complex coefficients of the
asymptotic series of $z$: \ba \label{z-k-coeff} z_{k}&=&
\frac{1}{\rho} \left[
\mu\sum_{m=0}^{k-1}\Delta_{k-m}\biggl(\sum_{j=0}^{m}
z_{m-j}z_{j}\biggr)-2(1+\rho)\sum_{m=1}^{k}
b_{m}z_{k-m}\right]\nonumber\\
&+&\frac{1}{\rho} \left[
\mu^*\Delta^{*}_{k}+i\,\frac{k-1}{2}z_{k-1}\right] \, . \ea For the
first coefficient, we get
$z_{1}=\mu^*\Delta^{*}_{1}/\rho=-\mu^*/(4\rho)$. Thus, for
sufficiently large values of $x$, the differential equation
(\ref{exact-z-eq}) should admit a solution $z$ such that\footnote{A
function $f(x)$ is $o(1/x^k)$ at infinity if
$\lim_{x\to\infty}x^kf(x)=0$.} \be \label{z-asymp}
z(x)=-\biggl(\frac{\mu^*}{4\rho}\biggr)\,
\frac{1}{x}+o\biggl(\frac{1}{x}\biggr) \, . \ee As we will see, this
result suffices to prove the square summability of the coefficients
$d_m$. Of course, the above behavior is not allowed for $z$ when
$2b_0\lambda-1=\rho$ vanishes, as it is the case for the Hamiltonian
$H$. This explains the break down of unitarity for that specific
case.

\subsection{Unitarity of the Modified Dynamics}
\label{sec-unitpert-ham}

Let us finally show that the deduced asymptotic behavior for $z(x)$,
together with relation (\ref{sqdm}), guarantee the square
summability of the sequence $\{d_m(t)\}$ for all fixed, strictly
positive values of the time coordinate $t$. From Eq.
(\ref{z-asymp}), we see that, at infinity, $\lim_{x\to
\infty}|xz+\mu^*/(4\rho)|=0$. So, given any constant $\varepsilon>0$
there exists a positive number $x_{0}(\varepsilon)$ such that $|
xz+\mu^*/(4\rho)| < \varepsilon$ for all $x>x_{0}(\varepsilon)$. One
can see that this inequality implies that, for all
$x>x_{0}(\varepsilon)$,\be |z|^{2} <
\frac{R(\varepsilon,\rho,\mu)}{x^{2}}\,,\quad
R(\varepsilon,\rho,\mu):=2\varepsilon^{2}+\frac{\varepsilon|\mu |}
{|\rho|} + \frac{|\mu|^2}{16\rho^{2}}\,.\ee

Let us now choose a number $\tilde{x}_{0}$ in the interval
$\big(x_{0}(\epsilon),\infty\big)$ such that
$\eta_{0}:=R(\varepsilon,\rho,\mu)/\tilde{x}_{0}^{2}<1$, which is
clearly always possible. Then, for all $x>\tilde{x}_{0}$, \be| z
|^{2} < \frac{R(\varepsilon,\rho,\mu)}{x^{2}}<\eta_{0}<1\,.\ee
Therefore, for all $x>\tilde{x}_{0}$ we have \be
\label{inequalities-d} \frac{| z |^{2}}{1-| z |^{2}}<\frac{| z
|^{2}}{1-\eta_{0}}<\frac{R_{0}}{x^{2}} \, , \quad
R_{0}:=\frac{R(\varepsilon , \rho,\mu)}{1-\eta_{0}}\,.\ee Notice
that $R_0$ is a finite and strictly positive constant. Employing Eq.
(\ref{sqdm}) and this inequality, we obtain that for all
$m>M_{0}:={\rm int}\{\tilde{x}_{0}/t\}+1$ (where ${\rm int}\{x\}$ is
the integer part of $x$) \be
|d(mt)|^{2}<\frac{R_{0}}{(mt)^{2}}\,.\ee Therefore, we conclude
that, for every fixed $t>0$, \ba \label{d-square-sum}
\sum_{m\in\N,\,m\neq 0}^{\infty}| d(mt)|^{2}&<& \sum_{m=1}^{M_{0}}|
d(mt)|^{2}+\frac{R_{0}}{t^{2}}\sum_{m=M_{0}+1}^{\infty}\frac{1}
{m^{2}}\nonumber\\ &<& \sum_{m=1}^{M_{0}}|
d(mt)|^{2}+\frac{\pi^{2}R_{0}}{6t^{2}}<\infty\, . \ea Thus, $d_m$ is
square summable, and the dynamics generated by $\tilde{H}$ in
$E_{B}$ is unitarily implementable. The proof, which makes use of
Eq. (\ref{z-asymp}), fails when $\rho=0$ and, in particular, for the
Hamiltonian $H$. We hence see that unitary implementability is
extremely sensitive to the value of $\rho$. For instance, for every
nonvanishing real $\rho$ in any neighborhood of zero (in fact, for
all $\rho \in \R-\{0\}$),
$\tilde{H}_{m}[{\cal{A}}]=H_{m}[{\cal{A}}]+2m\rho
b(x)(A_{m}^{*}A_{m}+A_{-m}^{*}A_{-m})$ gives rise to a unitarily
implementable transformation.

On the other hand, it is worth emphasizing that, to prove the square
summability of $d_m$, we have not actually employed the existence of
an asymptotic series for a solution $z$ of Eq. (\ref{exact-z-eq}).
What we have used in fact is a weaker property, namely, the
existence of a solution with the behavior (\ref{z-asymp}). A
rigorous proof of this existence is given in the appendix.

In our analysis, we have assumed a specific form for the
Hamiltonian, in particular that $\lambda$ and $\mu$ are a real and a
complex constant. It is nonetheless possible to generalize our
discussion to other cases. Suppose, e.g., that the function $\lambda
b(x)$ is replaced by a new real function $\bar{b}(x)$ and that $\mu
a(x)$ is changed into
$\bar{a}(x):=\exp{[-2i\psi(x)]}\bar{\Delta}(x)$ (and similarly for
its complex conjugate), with $\psi$ and $\bar{\Delta}$ being a real
and a complex function, respectively. Let us then call
$\bar{\beta}(x):=2\bar{b}(x)-d\psi(x)/dx$ and assume that
$\bar{\beta}(x)$ and $\bar{\Delta}(x)$ admit asymptotic series at
infinity such that $\bar{\Delta}_0=0$. Defining now
$\bar{z}(x)=\exp{[-2i\psi(x)]}Y(x)$, it is straightforward to see
that the same line of reasoning presented in the previous subsection
leads to an equation analogous to Eq. (\ref{exact-z-eq}), but with
the replacements of $z$ by $\bar{z}$, $[2\lambda b-1]$ by
$\bar{\beta}$, and $\mu\Delta$ by $\bar{\Delta}$. The counterpart of
Eq. (\ref{z-o-coeff}) is then $\bar{z}_{0}\bar{\beta}_0=0$, which
implies that $\bar{z}_{0}$ vanishes unless so does $\bar{\beta}_0$,
which plays now the role of $\rho$. In addition, the analog of Eq.
(\ref{z-k-coeff}) is \ba \bar{z}_{k}&=&
\frac{1}{\bar{\beta\,}_{0}}\left[\sum_{m=0}^{k-1}\bar{\Delta}_{k-m}
\left(\sum_{j=0}^{m}\bar{z}_{m-j}\bar{z}_{j}\right)-\sum_{m=1}^{k}
\bar{\beta}_{m}\bar{z}_{k-m}
\right]\nonumber\\
&+&\frac{1}{\bar{\beta\,}_{0}}\left[
\bar{\Delta}^{*}_{k}+i\,\frac{k-1}{2}\bar{z}_{k-1}\right] . \ea
Therefore, we get $\bar{z}(x)=\bar{\Delta}^*_1/(\bar{\beta\,}_0x) +
o(1/x)$. Again, this asymptotic behavior suffices to guarantee the
square summability of the antilinear part of the map $M_{m}(t)$ for
all $t>0$ provided that $\bar{\beta}_0$ differs from zero.

As a final comment, let us consider the formal quantum expression
for $\tilde{H}$: \be \label{formal-expr} :\widehat{\tilde{H}\,}: =
:\widehat{H}: + 2\rho\!\sum_{m\in \N,\,m\neq
0}m\,b(mt)\left[\hat{A}_{m}^{\dagger}\hat{A}_{m}
+\hat{A}_{-m}^{\dagger}\hat{A}_{-m}\right]. \ee Because of the
unitary implementability of $M_{m}(t)$, we know that
(\ref{formal-expr}) generates the unitary evolution operator through
which the basic operators $\hat{B}_{n}$ and $\hat{B}_{n}^{\dagger}$
evolve (when these basic operators are represented as the
annihilation and creation operators on the Hilbert space ${\H}_{B}$
constructed from $E_{B}$ and its corresponding complex structure
$J_{B}$). On the other hand, even though the operator
(\ref{formal-expr}) generates a map which acts unitarily on the
Hilbert space ${\H}_{\bar{\G}}$ (constructed from $\bar{\G}$ and its
associated complex structure $J_{\bar{\G}}$), we know that this map
does not correspond to the actual time evolution of the basic
operators $\hat{B}_{n}$ and $\hat{B}_{n}^{\dagger}$ (now represented
as the annihilation and creation operators on ${\H}_{\bar{\G}}$).
The quantum generator comes from a phase space function which
certainly can be considered as close as one wants to $H$ (the
generator of the dynamics in $\bar{\G}$), but does not coincide with
it. Nevertheless, it is interesting to note that if $\rho$ is
regarded as a constant of quantum origin, e.g. by setting $\rho$
proportional to $\hbar$, then the classical limit of
$\widehat{\tilde{H}\,}$ would be just $H$. That is, in spite of the
lack of a unitary implementation for $H$, if we consider that the
modification of the Hamiltonian arises from a quantum correction,
then we will get a unitary map whose generator, in the naive limit
$\hbar \to 0$, provides the classical dynamics in $\bar{\G}$.


\section{Conclusions and further comments}
\label{sec-conlusions}

We have analyzed the impossibility of obtaining a unitary
implementation of the dynamics in the polarized Gowdy $T^3$ model
with the quantization put forward in Ref. \cite{pierri}, a problem
that has recently been pointed out in Refs. \cite{ccq-t3} and
\cite{torre-prd}. With this aim, we have first presented a complete
derivation of the model starting with general relativity and
introducing a symmetry-reduction and gauge-fixing procedure.
Employing then a time-dependent map from the covariant phase space
to the canonical phase space of the system, we have been able to
reformulate the issue of unitary implementability of the evolution
as a question about the square summability of the antilinear part of
such a map. In this process, it is important to realize that the
total Hamiltonian in the covariant phase space vanishes, whereas the
considered map includes in an explicit manner all of the time
variation of the system. Exploiting this reformulation of the
unitarity problem, we have considered (certain types of)
modifications to the dynamics and analyzed whether the symplectic
maps associated with them are unitarily implementable. In this way,
we have traced back the failure of unitarity to the presence of some
specific contributions in the Hamiltonian that generates the
dynamics. In addition, we have seen that negligibly small
modifications of these contributions suffice to restore unitarity.
In the rest of the section, we present some comments about the main
results of the work.

In our analysis, two facts have played a particularly relevant role.
Firstly, as we have noticed, there is some kind of universality in
the behavior of the Bogoliubov coefficients. This has allowed us to
consider just one equation [namely Eq. (\ref{exact-z-eq})] in order
to examine the square summability of these coefficients, rather than
investigating an infinite number of differential equations, one for
each mode. Secondly, to know whether the modified Hamiltonian is
unitarily implementable, instead of solving the universal equation
(\ref{exact-z-eq}), it actually suffices to study the leading term
of the function $z$ in the asymptotic limit of large values of its
argument. In this sense, one does not need to explicitly integrate
the dynamical equations.

On the other hand, we note that solving Eq. (\ref{cond2}) amounts to
``diagonalizing'' the total Hamiltonian by means of a time-dependent
canonical transformation, namely, to requiring that the terms
proportional to $B_{m}B_{-m}$ and $B_{m}^{*}B^{*}_{-m}$ vanish in
the phase space function
$\tilde{H}_2[{\cal{B}}]:=\tilde{H}[{\cal{B}}]-
\partial_{t}{\cal{F}}[{\cal{B}}]$. Had we ignored the term containing
the time
derivative of the Bogoliubov coefficients, we would have obtained
from Eq. (\ref{cond2}) an algebraic quadratic equation for the ratio
$Y(x)$ of the coefficients that leads to an instantaneous
diagonalization (i.e., at a fixed instant of time) of the
Hamiltonian $\tilde{H}$. In fact, for $\lambda =\mu=1$ so that
$\tilde{H}$ reduces to $H$, one can see that using this algebraic
equation and the relation $| c |^{2} =1+| d |^{2}$, it is possible
to recover the instantaneous diagonalization given in Ref.
\cite{ccq-t3}. In addition, we emphasize that diagonalizing the
Hamiltonian is equivalent to the resolution of the dynamics. Indeed,
as we have seen, if we solve the universal equation
(\ref{exact-z-eq}), which is equivalent to the diagonalization
condition (\ref{cond2}), then $c(x)$ can be found by simple
integration of the first-order differential equation (\ref{eqc}),
whereas $d(x)$ is determined as $d(x)=\exp{(2ix)}\,z(x)c(x)$.

Our discussion can be extended to Hamiltonians for which the
functions $\mu a(x)$ and $\lambda b(x)$ in Eq. (\ref{gene-der-imp})
are replaced by more general functions $\bar{a}(x)$ and
$\bar{b}(x)$. We have seen that this is the case at least if
$\bar{a}(x)$ is of the form $\exp{[-2i\psi(x)}]\bar{\Delta}(x)$ with
$\psi(x)$ real, and $\bar{\Delta}(x)$ and (the real function)
$\bar{\beta}(x):=2\bar{b}(x)-d\psi(x)/dx$ admit asymptotic series
with a vanishing coefficient $\bar{\Delta}_0$. More precisely, we
have proved that the dynamics generated by those Hamiltonians can be
implemented as a unitary transformation as far as the coefficient
$\bar{\beta}_0$ differs from zero. In this sense, our study provides
a general treatment for Hamiltonians of the form
(\ref{gene-der-imp}), quadratic in the coordinates ${\cal A}_{m}$.

We have proved that the obstructions to unitarity are extremely
sensitive to perturbations of the Hamiltonian. In the quantization
performed in Ref. \cite{pierri}, one can in fact approximate the
dynamics as much as desired by means of unitary transformations.
Thus, although a thorough study of the influence of the choice of
deparametrization and complex structure must be performed in order
to elucidate if the failure of unitarity is a phenomenon inherent to
the polarized Gowdy $T^{3}$ model, our result suggests that it may
be actually possible to restore unitarity by considering a different
choice.

Finally, employing Eq. (\ref{abform}) and the asymptotic expansions
of the Hankel functions, it is not difficult to check that, for
large values of $t$, the $m$-th contribution to the total
Hamiltonian of the polarized Gowdy $T^3$ model [see Eq.
(\ref{gene-der-imp}) with $\lambda=1$] becomes
$H_{m}[{\cal{A}}]\approx m(A_{m}^{*}A_{m}+A_{-m}^{*}A_{-m})$. So,
the Hamiltonian for asymptotically large values of $t$ is given by
$H[{\cal{A}}]\approx \sum_{n\in \Z-\{0\}}| n | \,A_{n}^{*}A_{n}$,
which is a combination of harmonic oscillators. Therefore, we could
approximate the dynamics in an unitary way by means of some modified
Hamiltonian $\tilde{H}$ and, for large values of $t$, by a sum of
harmonic oscillators. Actually, corrections to the dynamics are not
unexpected, e.g., from a quantization in the framework of loop
quantum gravity. It might happen that some kind of (fictitious)
effective dynamics could take place as a result of the smearing of
the initial singularity by quantum effects (like it occurs, in fact,
for a flat isotropic universe, where the cut-off for curvatures
provides us with an effective Friedmann equation \cite{bojowald}).
This effective dynamics could correspond to a perturbation of the
evolution generated by $H$ which might be unitarily implementable.


\section*{Acknowledgments}

The authors are grateful to A. Corichi, J.F. Barbero G., and E.J.S.
Villase$\tilde{\rm n}$or for helpful comments. J. Cortez was funded
by the Spanish Ministry of Education and Science (MEC), No./Ref.
SB2003-0168. This work was supported by funds provided by the
Spanish MEC Projects No. FIS2004-01912, No. BFM2002-04031-C02, and
No. HP2003-0140.


\begin{appendix}
\section{Proof of the asymptotic behavior}
\label{sec-proof}

In this appendix, we want to prove that Eq. (\ref{exact-z-eq})
admits one solution which, at infinity, has the asymptotic behavior
(\ref{z-asymp}), provided that $\rho\neq 0$. Let us start by
defining a new function $w(x)$ by means of the relation  \be
\label{defw} z(x)=-\frac{\mu^*}{4\rho x}+\frac{w(x)}{x}.\ee
Substituting this expression in Eq. (\ref{exact-z-eq}) we obtain an
equivalent nonlinear differential equation of the Riccati type\be
\label{weq}\frac{dw}{dx}(x)=w(x)\beta(x)-2i\mu\frac{\Delta(x)}{x}
\,w^2(x)+\alpha(x)\,,\ee where \ba\label{alph-bet}
\beta(x)&:=&2i\left[2\lambda
b(x)-1\right]+\frac{1}{x}+i\,\frac{|\mu|^2}
{\rho}\frac{\Delta(x)}{x}\,,\nonumber\\
\alpha(x)&:=&-2i\mu^*x\left[\Delta^*(x)-\Delta_1^*\right]
-i\,\frac{\mu^*\lambda}{\rho}\left[b(x)-b_0\right]\nonumber\\
&-&\frac{\mu^*}{4\rho
x}-i\,\frac{\mu^*|\mu|^2}{8\rho^2}\frac{\Delta(x)}{x}\,.\ea The
constants $\Delta_1$ and $b_0$ are given in Eq. (\ref{coefbD}) and
we have used $\lambda=1+\rho$.

In order to arrive at the desired result about the asymptotic
behavior of $z(x)$, we only have to demonstrate that Eq. (\ref{weq})
admits a solution that tends to zero at infinity.

Employing the asymptotic expansions (\ref{asympt-b-delta-2}) of the
functions $\Delta(x)$ and $b(x)$, recalling that $\Delta_0=b_1=0$,
and making use of Eq. (\ref{asympt-b-delta-1}) to compute the
coefficient $\Delta_2=i/16$, one can rewrite\ba\label{orderalph-bet}
\beta(x)&=&2i\rho+\frac{1}{x}+\tilde{\beta}(x)\,,\nonumber\\
\alpha(x)&=&\left(\frac{1}{2}-\frac{1}{\rho}\right)\frac{\mu^*}{4x}
+\tilde{\alpha}(x)\,,\ea with $\tilde{\alpha}(x)$ and
$\tilde{\beta}(x)$ being $O(1/x^2)$ at infinity [we say that a
function $f(x)$ is $O(1/x^n)$ at infinity if $|x^nf(x)|$ admits a
finite limit when $x\rightarrow\infty$]. Explicitly, these functions
are \ba
\tilde{\beta}(x)&:=&4i\lambda\left[b(x)-b_0\right]+i\frac{|\mu|^2}
{\rho}\frac{\Delta(x)}{x}\,,\nonumber\\
\tilde{\alpha}(x)&:=&-2i\mu^*x\left[\Delta^*(x)-\Delta_1^*-
\frac{\Delta_2^*}{x}\right]-i\,\frac{\mu^*\lambda}{\rho}
\left[b(x)-b_0\right]\nonumber\\
&-&i\, \frac{\mu^*|\mu|^2}{8\rho^2}\frac{\Delta(x)}{x}\,.\ea On the
other hand, the function $\Delta(x)/x$ that multiplies $w^2(x)$ in
Eq. (\ref{weq}) is also $O(1/x^2)$ asymptotically. For solutions
$w(x)$ that are small at infinity, we then expect the quadratic term
in our Riccati equation to be negligible. We will hence approximate
our equation by a linear one that can be explicitly solved, find for
it a solution that tends to zero at infinity, and prove that, for
that solution, the removed quadratic term can in fact be neglected
in the original differential equation.

For the linear differential equation \be\label{lineq}
\frac{dw_l}{dx}(x)=w_l(x)\beta(x)+\alpha(x)\ee all solutions can be
constructed starting with those of the associated homogeneous
equation. Using Eq. (\ref{orderalph-bet}), the homogeneous solutions
can be found to be proportional to \be w_l^h(x)=x\exp{\left[2i\rho
x+\int^x_{\infty}d\bar{x}\;\tilde{\beta}(\bar{x})\right]}\,.\ee Note
that the asymptotic behavior of $\tilde{\beta}$ guarantees that the
integral that appears in this expression is well-defined. Solutions
to Eq. (\ref{lineq}) [modulo the possible addition of $w_l^h(x)$
times a complex constant] are then of the form \be\label{solin}
w_l(x|x_0)=w_l^h(x)\int^x_{x_0}d\bar{x}
\frac{\alpha(\bar{x})}{w_l^h(\bar{x})}\,,\ee where $x_0$ is a
constant. A convenient integration by parts leads then to \ba
w_l(x|x_0)&=&w_l^h(x)\!\int^x_{x_0}\!d\bar{x}\,
\frac{\bar{x}}{w_l^h(\bar{x})}
\frac{d\;}{d\bar{x}}\left[\frac{\alpha(\bar{x})}
{\left\{2i\rho+\tilde{\beta}(\bar{x})\right\}\bar{x}}\right]
\nonumber\\
&+&w_l^h(x)\left[\left.\frac{\alpha(\bar{x})}{w_l^h
(\bar{x})\left\{2i\rho+\tilde{\beta}(\bar{x})\right\}}
\right|_x^{x_0} \right].\ea Remembering that $\alpha(x)$,
$\tilde{\beta}(x)$, and $w_l^h(x)/x$ are respectively $O(1/x)$,
$O(1/x^2)$ and $O(1)$ at infinity, it is possible to see that the
integrand in the above expression is $O(1/x^3)$. Therefore, the
integral converges when $x_0$ tends to infinity.

In that limit, one gets the particular solution \ba\label{wlinpart}
w_l^P(x)&=&w_l^h(x)\!\int^x_{\infty}\!d\bar{x}\,
\frac{\bar{x}}{w_l^h(\bar{x})}
\frac{d\;}{d\bar{x}}\left[\frac{\alpha(\bar{x})}
{\left\{2i\rho+\tilde{\beta}(\bar{x})\right\}\bar{x}}\right]
\nonumber\\
&-&\frac{\alpha(x)}{2i\rho+\tilde{\beta}(x)}.\ea One can see (e.g.
using L'H\^{o}pital's rule for the term containing the integral)
that this solution tends to zero when $x\rightarrow\infty$.
Furthermore, repeating the explained procedure of integration by
parts, one can show that the total contribution to $w_l^P(x)$ coming
from the factor that includes the integral is $O(1/x^2)$. Using Eq.
(\ref{orderalph-bet}), one then concludes that the asymptotic
behavior of $w_l^P(x)$ is \be
w_l^P(x)=-\left(\frac{1}{2}-\frac{1}{\rho}\right)\frac{\mu^*}{8i\rho
x}+o\left(\frac{1}{x}\right)\,.\ee It is now a simple exercise to
check that, in the Riccati equation (\ref{weq}), the quadratic term
is $o(1/x^3)$ at infinity for the solution $w_l^P(x)$, which is in
fact negligible when compared with the rest of terms in the equation
(in particular with $dw/dx$), which are at least of order $1/x^2$.
This concludes our proof.
\end{appendix}


\end{document}